# Interface Matching and Combining Techniques for Services Integration

Frédéric Le Mouël, Noha Ibrahim, and Stéphane Frénot

*Abstract*—The development of many highly dynamic environments, like pervasive environments, introduces the possibility to use geographically closely-related services. Dynamically integrating and unintegrating these services in running applications is a key challenge for this use. In this article, we classify service integration issues according to interfaces exported by services and internal combining techniques. We also propose a contextual integration service, IntegServ, and an interface, Integrable, for developing services.

## I. INTRODUCTION

The development of many highly dynamic environments, like pervasive environment [22][23] or large-scale grid [25], offers the possibility for applications to use geographically-closed and closely-related services of the environment. Indeed, applications would like, whenever it is possible or needed, to integrate services provided by the local environment. In particular, if no single service can satisfy the functionality required by an application, a combination of existing services should be realized and integrated in order to fulfill the request.

Nowadays, taking into account and integrating new services in applications is not possible without the need of complicated processes [7]. Hence, we propose in this paper a developing framework and a run-time environment with automatic, smart and customizable integration of services. Our system, called ANIS - *Automated Negotiated Integration System*, is generic from application development point of view with an `Integrable` interface, and from system point of view while implementing different techniques such as composition, weaving, parameterization or deployment.

This paper is organized as follows. In section II, we propose a classification of service integration issues according to interfaces exported and internal combining techniques. In section III, we detail our ANIS system, especially the IntegServ service, `Integrable` interface and the integration by composition. Section IV presents related integration works in different domains. Finally we conclude and give future research works.

## II. SERVICE INTEGRATION CLASSIFICATION

Integration is the process of incorporating a service or a set of services so that they can work together and provide a new service [3]. Before to integrate services, we firstly consider in subsection A the model of our service. Then we detail impact of the integration on the different parts of a service. Subsection B tackles external concerns, especially exported interface matching. Subsection C undertakes internal concerns with the different combing techniques.

### A. Service Model

As shown in the figure 1, our service is composed of four parts:
o *Interfaces*: an interface specifies methods that can be performed on the service. Service's interfaces are public and published for an external use. A service can hold two kinds of interfaces: *functional interfaces* defining the functional behavior of the service (e.g. for a video streaming service, a functional interface can allow to specify frame's size, frame's rate, etc.) and *management interfaces* defining the way to manage this service (e.g. a life-cycle interface can allow to specify when to start/stop a service).
o *Bindings*: a service can provide and/or require functionalities from other services. Bindings express these run-time dependencies (e.g. if a video streaming service requires a QoS communication interface, at one moment, it can bind to a H.323 service, and at some later moment, to a SIP/RTP service).
o *Objects*: objects realize the functionality expected from the service (e.g. in our video streaming service, objects multiplex/demultiplex, order/reorder video and audio frames, etc.)
o *Context*: context models the service's run-time environment. Two services can have same interfaces, bindings and objects but can be a different point of their execution, e.g. one is started and the other is stopped. A context can maintain internal information, such as the current service's parametrization, the running state, or the user's profiling and customization. Or it cal also maintain external information, such as locations, other environment's available services, etc.

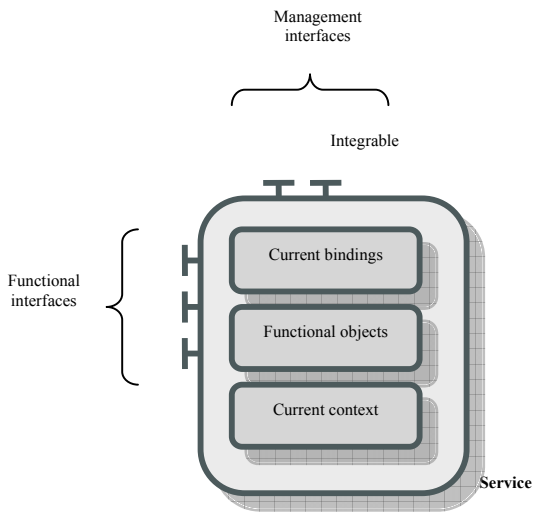

Fig. 1. Service's structure

Our service model is independent of any implementations and can be applied to EJBs [14], CORBA Components [27], Fractal components [2], OSGi bundles/services [16] or Web services [9].

Technically, interfaces can be expressed in language-native way, such as Java interfaces or by using an Interface Description Language (IDL), such as in CORBA [15]. Bindings can be expressed by an Architecture Description Language (ADL) [20]. Objects implementations are language-dependant and results from the instantiation of classes. For the context, a great variety of techniques can be used according to the reached goal. For instance, for keeping an internal state, serialization can be used to save parameters.

*B. Interface Matching*

The first step in an integration process is to find common functionalities present in services we want to combine.

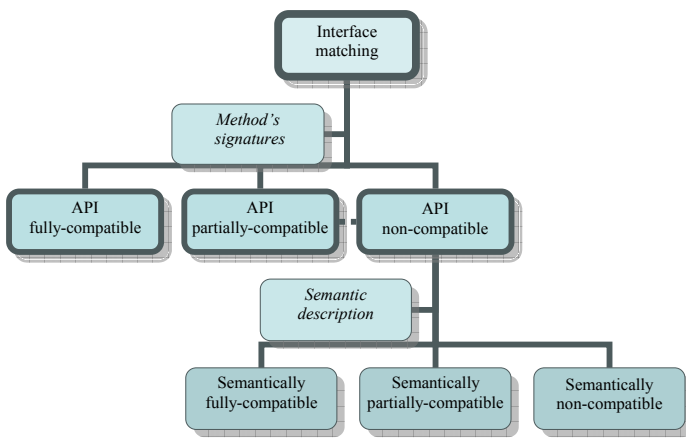

Fig. 2. Interface matching ontology

As interfaces are publicly published and define operations which can be performed on the service, this problem consists in an interface matching.

Figure 2 shows the compatibility possibilities according to two criteria. The first criterion we examine is the method's signature. It consists in comparing all method's names, method's input/output types. This comparison can result in:

o *API fully-compatible matching*: all method's signatures exactly match. Combining techniques can be applied.
o *API partially-compatible matching*: only a subset of method's signatures exactly matches. Combining techniques can be applied to this subset.
o *API non-compatible matching*: No method's signatures exactly match. No combining techniques can be applied.

Another criterion, the semantic description of interfaces, can be taken into account for interface matching. Systems do not always provide this information, but when present, such as with [24][26], it can parallelly be used with method's signature criterion and is especially useful when API is only partially or non compatible. This second criterion comparison can result in:

o *Semantically fully-compatible matching*: as methods semantically match, a process of transformation of method's name and method's input and output data can be applied (e.g. a proxy can redirect method's calls). After this transformation process, the combining techniques can be applied.
o *Semantically partially-compatible matching*: only a subset of methods semantically matches. Transformation techniques and then combining techniques can be applied to this subset.
o *Semantically non-compatible matching*: no integration can be done.

*C. Combining Techniques*

Now we have found common functionalities in our services, the second step consists in combining services to provide a new one. This combining involves internal parts of services, i.e. bindings, objects and context.

Integrating a service must be locally possible but, as we focus on integration in highly dynamic environments, one key challenge is also to allow distant services integration. Figure 3 shows these two possibilities:

o *Local combining techniques*: services hosted on the same machine can be simply or optimally combined.

  *Simple combining* consists in adding all functionalities of services in a new one (c.f. figure 4). Composition allows to simply combine services by connecting interfaces and updating bindings; the new service just redirects method's calls. Weaving also allows to simply combine services by generating new objects and new context; new objects results from interlacing instructions inside of methods.

  *Optimized combining* consists in selecting appropriate functionalities of each service for the new one (c.f. figure 5). These optimizations can consist in removing some methods, for instance, if they are redundant or non-useful (methods are not required by external services). Another optimization is to choose appropriate methods depending of the context. For example, if the run-time context is a PDA or a memory card, only memory-limited methods can be chosen.

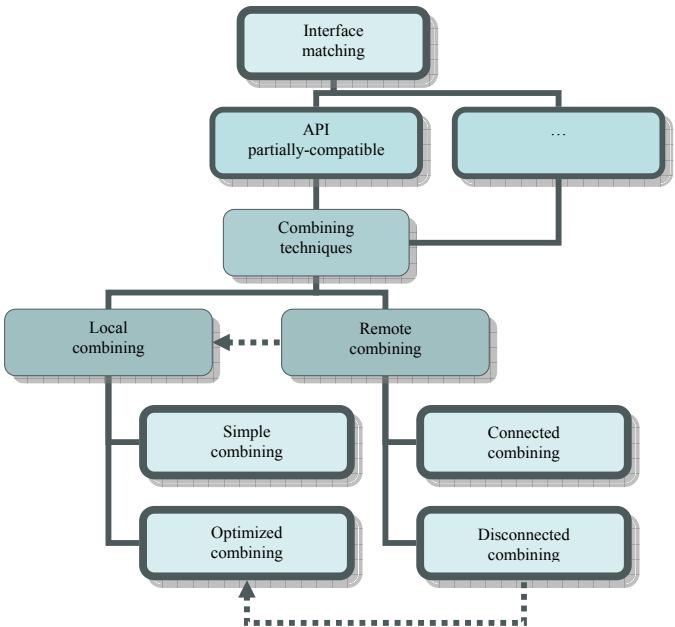

Fig. 3. Combining techniques ontology

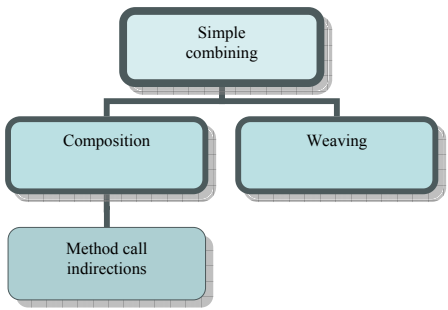

Fig. 4. Simple combining implementations

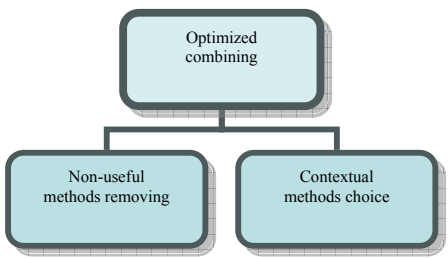

Fig. 5. Optimized combining implementations

o *Remote combining techniques*: services hosted on distant machine can be combined in a connected or disconnected way.

*Connected combining* consists in adding communication objects to service's objects. For instance, by using added stubs and skeletons, local method's calls are then transformed to remote calls, such as Remote Procedure Call (RPC), Remote Method Invocation (RMI) or event-based calls.

*Disconnected combining* consists in adding communication objects which anticipate and palliate disconnections. Different techniques exist such as proxying or caching. These techniques can solve consistency problems (method's calls ordering, etc) or can choose contextual-appropriate heuristics (semantic choice of important calls).

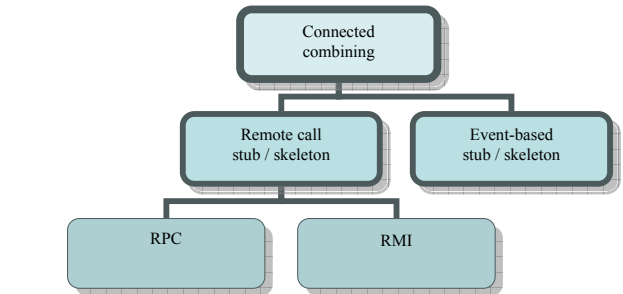

Fig. 6. Connected combining implementations

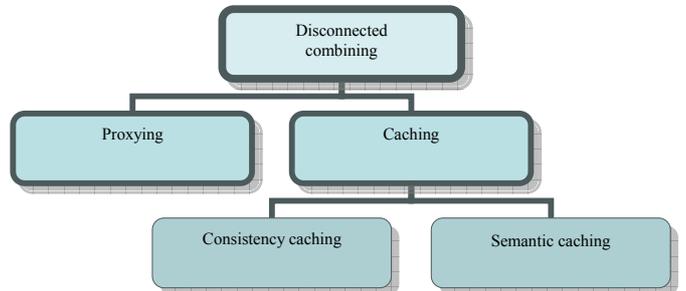

Fig. 7. Disconnected combining implementations

These techniques require to add and deploy additional objects to services. For these additional objects, local combining techniques can also be applied. For example, we remotely integrate the service 1 on host A with the service 2 on host B. Consistency caches are added on host A and B. The service 1 can simply composed with its cache on host A, while service 2 can decide to optimize non-useful methods with its cache on host B.

III. ANIS: AUTOMATED NEGOTIATED INTEGRATION SYSTEM

To tackle the different concepts introduced in section II, we implement a developing framework and a run-time environment with automatic, smart and customizable integration of services. Our system, called ANIS - *Automated Negotiated Integration System*, proposes (i) an `Integrable` interface for developing services (subsection B), (ii) an IntegServ service which realizes the integration at run-time (subsection A) and (iii) a toolkit with implementations of the different techniques presented in section II (subsection C).

### A. Architecture

One key part of our system is the IntegServ service. This service is called by all services to realize the integration. As the figure 8 shows, this service is itself the result of the integration of four other services:

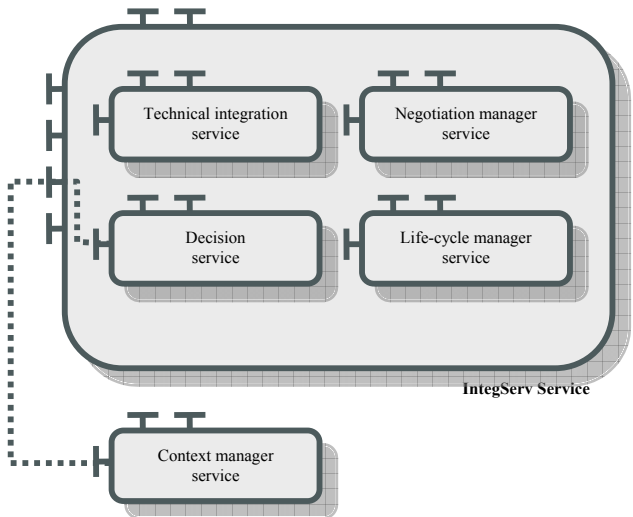

Fig. 8. IntegServ service architecture

o *The decision service*: the capacity of decision of our IntegServ service is provided by the decision service. This service can take decisions and adapt automatically to the variations of context. The decision service uses strategies to decide the proper integration to apply. These strategies should be made based on up-to-date information due to the highly dynamic nature of the environment.

o *The negotiation manager service*: this service offers the possibility to services to come to an agreement on terms and conditions of their integration. It occurs when certain services requested for integration are not available, or when context changes and requires a re-analysis of integration. It offers an alternative to the primary integration decided before.

o *The life cycle manager service*: an integration can have, from its creation, a life time, known and managed by the life cycle manager. Once this time expires, the life cycle manager service informs the decision service which unintegrates this integration. Changes in the context can also have impacts on the life cycle of an integration. For instance, when one pre-integrated service leaves the context, the context manager service informs the decision service about it. The decision service decides about a new (un)(re)integration and informs the life cycle manager service of the update.

o *The technical integration service*: the technical integration service is the service which allows applying the different combining techniques. It is part of the basic services and carries out the orders of the decision service. The different techniques can be applied one before the other and/or combined.

### B. *Integrable Interface*

To let developers easily implement integrable services, we define the `Integrable` interface. It provides three methods allowing to manage the integration in a service, of a service or group of services: `integrate`, `unintegrate` and `getIntegratedServices` methods.

```
public interface Integrable {

  void integrate(Collection serviceSet)
  throws IntegrationException;

  void unintegrate(Collection serviceSet)
  throws UnIntegrationException;

  Collection getIntegratedServices();

}
```

`integrate` methods allows integrating a set of services within the current service. The IntegServ service is called by this method and decides which technical combining techniques to apply (weaving, composition, optimized or not, etc) (c.f. section II.C). In case the integration is not possible an `IntegrationException` exception is raised. This case of error can appear if for instance we undertake a weaving between objects unweavable.

`unintegrate` method allows to cancel integration of a group of services beforehand integrated. It guarantees the reversibility of integration. In case the service to disintegrate is being used in the context or is not available, a `UnIntegrationException` exception is raised.

`getIntegratedServices` method returns all services having already been integrated into the current service.

### C. *Integration by Local and Remote Composition in an OSGi Framework*

We implement our developing framework on an OSGi platform and enrich it with two combining techniques: local and remote composition.

We apply our service model to OSGi's bundle and service: Interfaces are Java interfaces; objects are Java runtime objects instantiation of classes started by an `Activator`; bindings are modeled by the `manifest.mf` file.

As shown in figure 9, Service1 hosted on machine A and Service2 hosted on machine B implement the `displayHelloWorld ()` method. We integrate service2 to service1 by calling the integrate method of service1:

```
Service1.integrate(new HashSet(Service2));
```

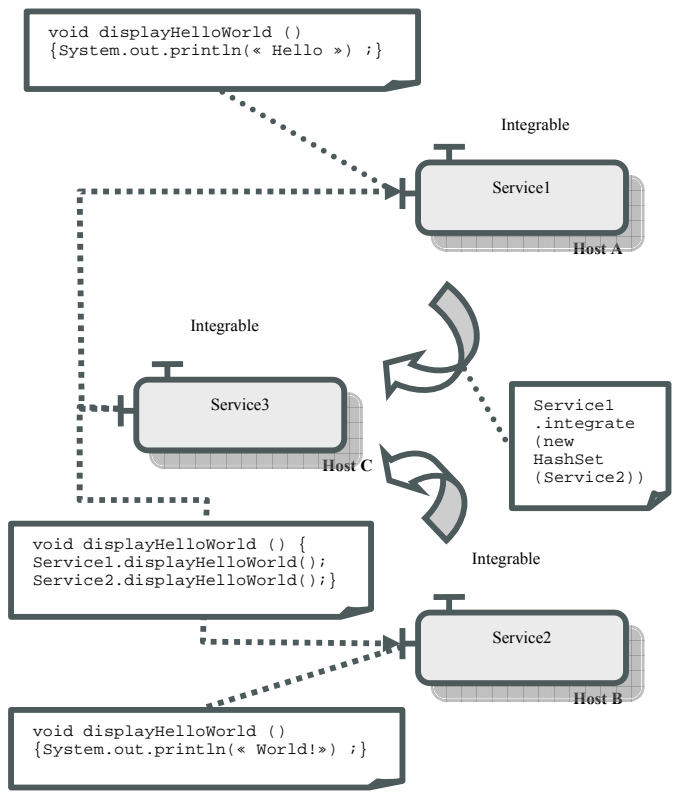

Fig. 9. Example of services integration by composition

This integrate call uses the IntegServ that creates by composition a new service Service3 on host C. This service offers the same method than Service1 and Service2 and remotely redirect the `displayHelloWorld()` call in sequence to `Service1.displayHelloWorld()` and `Service2.displayHelloWorld()`. This sequence ordering depends on the combining techniques applied.

## IV. Related Work

Three major domains of object oriented programming lean over the concept of integration: Component-Based Software Engineering (CBSE) [8], Aspect-Oriented Programming (AOP) [9] and Service-Oriented Programming (SOP) [1]. Each of these domains has several definitions and techniques of integration according to the different existent platforms. Highly dynamic environments become more and more a target domain for this type of programming and the integration of services takes a new sense.

Different types of models based on components as EJBs [14], CORBA Component Model [27], Fractal [2] and Web services [9] allow the interaction between distant components. The integration of components in these different models is often reduced to the deployment and/or the parameterization of these components. However, these integrations are not perfectly adapted to highly dynamic environments and they do not take into account the change of context at the time of execution or the deployment of components. In these models, the definition of new components is rather difficult during execution, so the integration of components is often predefined beforehand. The development of pervasive environments throws a certain number of new challenges for component programming based, especially concerning taking into account mobility, context awareness and adaptability. Molène and AeDEn [13] projects offer an approach consisting of an adaptive distribution of applications allowing using resources of the environment dynamically to palliate the insufficiency of the resources of the mobile. AURA [6] project proposes a model of programming based on task. In this model, tasks are seen as being a composition of several components. AURA interprets the physical context of the user and can thus discover and compose components to fulfill a task.

Aspect-Oriented programming allows to establish transverse concerns (aspects) independent ones of the others and to combine them (the weaving) later to produce final application. AspectJ [11], Fac [17] and [5] are models based on aspect, applying the weaving of aspect as method of integration. Recent works were fulfilled on adaptation seen as an aspect in pervasive environments [19]. By using the aspects of AspectJ, the system modularizes three essential faces of adaptation in pervasive environments: management of the devices present in context, management of their contents, as well as the adaptation of devices to the change of context.

In the terminology of Service-Oriented programming, the integration of service is often reduced to a composition of service. Nowadays, researches aim at developing an architecture which allows the composition of service by using a logical reasoning given by the languages of description of service as DAML [24], Universal Description, Discovery and Integration (UDDI) and Web Service Description Language (WSDL) [26]. These languages define standard ways for service discovery, description and invocation (message passing). SWORD [18] is a developer toolkit for building composite web service. It does not deploy the emerging service description standards such as WSDL and DAML-S, instead, it uses rule-based plan generation, and it specifies the web services by using Entity-Relation model. Many current service composition platforms have been designed with the inherent assumption that the services are resident in the fixed network infrastructure and running on a relatively stable platform. Few have tried to consider alternate design approaches of service composition systems for highly dynamic environments. A distributed broker-based service composition protocol for pervasive environments [4] proposes a model adapted for pervasive computing, but it focuses only on the composition aspect of integration. For each composite request, the protocol elects a Broker from within a set of nodes. The request source delegates the responsibility of composition (i.e. discovery, integration and execution) to the elected broker. The main protocol, based on the composition and the integration, is seen to be a part of the protocol of composition. Scooby [21] a middleware for service composition in pervasive computing, proposes a system which provides a solution based on the use of binding variables utilizing late and lazy dynamic binding, along with the supporting service composition language in which users can formally specify their policies based on event notification messaging system.

## V. CONCLUSION AND FUTURE WORKS

In this article, we focus on service integration, especially in highly dynamic environments. We classify existing integration solutions according interface matching and combining techniques. We also propose a developing framework and a run-time environment with automatic, smart and customizable integration of services. Our ANIS - *Automated Negotiated Integration System* allows to easily develop and integrate services by using the `Integrable` interface. The IntegServ service provides technical integration service, negotiation service, decision service and life cycle manager.

In the future, we aim at finishing the development of our system under OSGi and publishing our services as UPnP services. We are also working on adding a semantic description of our services so as to enrich the negotiation and decision services.